\documentclass[aps,pra,twocolumn,superscriptaddress,floatfix]{revtex4-1}
\usepackage{hyperref}
\usepackage{graphicx,amsmath, amssymb, amsthm, subfigure}
\usepackage{xcolor}
\usepackage{braket}

\begin{document}

\title{Inverted Ladder Type Optical Excitation of Potassium Rydberg States with Hot and Cold Ensembles}
\author{Tzu-Ling Chen}
\email[]{Now with Division of Chemistry and Chemical Engineering, California Institute of Technology, Pasadena, California 91125, USA}
\affiliation{Department of Physics, National Tsing Hua University, Hsinchu 30013, Taiwan}
\affiliation{Center for Quantum Technology, National Tsing Hua University, Hsinchu, 30013, Taiwan}
\author{Shao-Yu Chang}
\affiliation{Department of Physics, National Tsing Hua University, Hsinchu 30013, Taiwan}
\affiliation{Center for Quantum Technology, National Tsing Hua University, Hsinchu, 30013, Taiwan}
\author{Yi-Jan Huang}
\affiliation{Institute of Photonics Technologies, National Tsing Hua University, Hsinchu 30013, Taiwan}
\author{Khemendra Shukla}
\affiliation{Department of Physics, National Tsing Hua University, Hsinchu 30013, Taiwan}
\affiliation{Center for Quantum Technology, National Tsing Hua University, Hsinchu, 30013, Taiwan}
\author{Yao-Chin Huang}
\email[]{Now with Institute of Photonics Technologies, National Tsing Hua University, Hsinchu 30013, Taiwan}
\affiliation{Department of Physics, National Tsing Hua University, Hsinchu 30013, Taiwan}
\affiliation{Center for Quantum Technology, National Tsing Hua University, Hsinchu, 30013, Taiwan}
\author{Te-Hwei Suen}
\affiliation{Department of Physics, National Tsing Hua University, Hsinchu 30013, Taiwan}
\affiliation{Center for Quantum Technology, National Tsing Hua University, Hsinchu, 30013, Taiwan}
\author{Tzu-Yung Kuan}
\affiliation{Department of Physics, National Tsing Hua University, Hsinchu 30013, Taiwan}
\affiliation{Center for Quantum Technology, National Tsing Hua University, Hsinchu, 30013, Taiwan}
\author{Jow-Tsong Shy}
\affiliation{Department of Physics, National Tsing Hua University, Hsinchu 30013, Taiwan}
\affiliation{Institute of Photonics Technologies, National Tsing Hua University, Hsinchu 30013, Taiwan}
\author{Yi-Wei Liu}
\email{ywliu@phys.nthu.edu.tw}
\affiliation{Department of Physics, National Tsing Hua University, Hsinchu 30013, Taiwan}
\affiliation{Center for Quantum Technology, National Tsing Hua University, Hsinchu, 30013, Taiwan}

\date{\today}

\begin{abstract}
We present experimental results on the sub-Doppler Rydberg spectroscopy of potassium in a hot cell and cold atoms, using two counter-propagating laser beams of 405~nm and 980~nm as a inverted
ladder-type excitation
configuration (4S$_{1/2}$-5P$_{3/2}$-$n$S$_{1/2}$ and $n$D$_{3/2,5/2}$). Such an inverted ladder-type scheme is predicted to be without sub-Doppler electromagnetically induced transparency (EIT) feature in a thermal ensemble under the weak-probe approximation. Instead, we utilized a strong probe field and successfully observed a transparency window with a width narrower than 50~MHz. Our all-order numerical simulation is in satisfactory agreement with the experimental results. This narrow linewidth allows us to measure the energy levels of the Rydberg levels from $n$=20-70 with improved accuracy. The deduced ionization energy agrees with the previous measurements. Furthermore, the two-photon Rydberg excitation scheme was applied to the cold ensembles to study the ground-state atoms population decrease in the MOT for various Rydberg states. Our experimental observations suggested two distinct regimes of the trap losses under different probe detuning conditions. While the far off-resonance case ($\Delta_p$$\gg$0) can be described by the picture of dressed atom, the on-resonance case ($\Delta_p$$\sim$0) reveals more interesting results. The higher Rydberg states suffer larger trap loss. Besides, even with similar level energies, the excitation to $n$D states result in faster escape of the ground-state atom from trap than nearby $n$S states.
     
\end{abstract}

\maketitle
\section{Introduction}
A Rydberg atom, which is an excited atom with one or more electrons in a high principal quantum number ($n$) state. Because of its extremely large polarizability, it has gained growing interest recently. Particularly, for quantum information processing, such a controllable large polarizability introduces many advantageous characteristics including strong dipole-dipole interactions that scale as $n^4$ and long radiative lifetimes that scale as $n^3$. The ability of controlling the dipole-dipole interactions in Rydberg systems makes neutral atomic qubits unfold extraordinary potential to compete with trapped ion qubits as building block of quantum gates \cite{PhysRevLett.123.170503, Wilk:2010il,Isenhower:2010db,Lukin:2001gg}. Meanwhile, the Rydberg dressed ensemble, also as a playground of collective phenomena, sheds new light on quantum simulation \cite{Bernien:2017bp,Zeiher:2016ku,Labuhn:2016dp,Dudin:2012hm} and enhances the optical non-linearity induced by two-photon atomic coherences for quantum optics applications\cite{Maxwell:2013gb,Peyronel:2012je,Pritchard:2010im,Mohapatra:2008hh,Liu:2014ct, Sheng:2017es}.

Since many works with optical trapped neutral atoms systems had contributed in improving the fidelity of qubit gate operations, one of the outstanding challenges in encoding qubits is the implementation of quantum non-destructive qubit state measurements without loss \cite{Saffman:2010ky}.
Besides using spatial localization to focus on one specific atom, using heteronuclear qubits is an improved method to mitigate crosstalk in neutral atom array experiments for the reason that they allow the state measurements possessing cross entanglement of two different atomic species located in the same trap, or nearby traps. The feasibility of this scheme has been discussed in \cite{Saffman:2016ea, Qian:2015jf, Beterov:2015ij}, and experimental research relevant to heteronuclear atom qubits has been reported in Rubidium (Rb) isotopes \cite{Zeng:2017dy}.
Since the heteronuclear atom qubits need the pre-preparation of ultracold alkali metal dimers. Only several combinations of them enable keeping stable under two-body molecular collisions \cite{zuchowski2010reactions}.
Aside from the suggested Rb-Cs configuration, potassium and rubidium (K-Rb), is another possible candidate for the heteronuclear Rydberg qubits, in which both species share a similar cooling laser system.
The two-body interaction in Rydberg-dressing schemes has been theoretically studied in \cite{Samboy:2017in}. Meanwhile, the Rydberg dressed fermionic isotope of potassium $^{40}K$ could provide richer collective phenomena and deeper insight for strongly correlated physics \cite{Mattioli:2013jl,Li:2015hl}.
Also, the potassium, which owes two isotopes $^{39,40}$K enable for dimer association, provides a wide range of scenarios for studying heteronuclear Rydberg qubits \cite{Grobner2016}.
However, the Rydberg excitation on potassium, which is essential for the interatomic interaction in future K-Rb heteronuclear Rydberg systems, was less studied and reported.

The species most commonly used for Rydberg experiments are alkali metals.
However, the work with K have scarcely been explored experimentally compared to which in Rb \cite{Li:19}. Thus, more comprehensively, we experimentally studied the optical excitation of K Rydberg states using both hot and cold atoms, together with numerical simulation comparison. We measured EIT spectra using two-photon excitation with an inverted ladder type configuration ($\lambda_c >\lambda_p$), where the wavelength of the probe field (405~nm) is much shorter than that of the coupling field (980~nm).
The inverted scheme is particularly suitable for the future realization of multiqubit gates \cite{PhysRevLett.123.170503}.
The coupling laser system at the wavelength can be benefited from the well-developed power amplifier with a power up to watt level in the near-infrared range \cite{Huang:2018cr}. Towards the heteronuclear Rydberg atoms interaction applications, such a high-power laser also acts as an optical dipole trap for the ultracold atomic ensembles of K and Rb.

Despite the inverted ladder type scheme in hot atomic ensembles was known to be lack of sub-Doppler feature, such as EIT, in the weak probe approximation using simple three-level model \cite{Urvoy:2013if}. However, the sub-Doppler feature was explicitly observed in our experiment. Thus, we perform theoretical simulation beyond the weak probe approximation to compare with the experimental observations.
Our excitation scheme was also performed with a cold ensemble in a magneto-optical trap (MOT) using the trap loss as detection signal. The two-photon excitation with the normal ladder scheme ($\lambda_c <\lambda_p$) in cold Rydberg ensemble was reported in \cite{Arias:2017ct}. The state-dependent trap loss was studied in our experiments to provide the information for the attempts at Rydberg dressing experiments. The interaction of cold Rydberg atoms has been realized with atom pairs \cite{PhysRevA.89.011402} and optical lattices \cite{Zeiher:2017iv,Zeiher:2016ku}, where atoms are orderly distributed. In contrast, in a homogeneous distributed atomic cloud, such as MOT or Bose-Einstein Condensate (BEC), an unexpectedly atom loss that has been observed in many experiments poses a challenging task, and the mechanism behind remains unclear. Only recently, A. D. Bounds \emph{et al.} \cite{Bounds:2018ei} reported a stable Rydberg-dressed MOT of Sr with a lifetime of several ms at a temperature $<1$~$\mu$K. In our experiment, we employed the steady state approach to show that the trap loss, induced by the interaction between the ground-state and Rydberg atoms, depends not only on the principle quantum number $n$, but also the orbit angular momentum $l$.

\section{Theoretical Model}\label{set:model}
Two-photon excitation is widely used to excite the atom to the Rydberg state in a three-level ladder-type system. It is particularly useful in those cases where one-photon excitation requires energy of deep UV wavelength or the excited Rydberg states has the same parity with the ground state.
The two-photon atomic coherence, resulted from a coherent interaction of the coupling and probe lasers with atoms, introduces a variety of interesting phenomenon in the ladder-type Rydberg atom, such as EIT \cite{Fleischhauer:2005ti}, electromagnetically induced absorption (EIA) \cite{Akulshin:1998wv}, and Autler-Townes splitting effect. 
The ladder-type EIT takes advantage of the dramatic changes in the optical properties to achieve high-resolution Rydberg spectroscopy and to demonstrate a direct non-destructive optical detection of highly excited Rydberg states \cite{Mohapatra:2007kb}. More complex phenomenon were also studied taking noncoherent effects into account \cite{Bhowmick:2016ima}.

\begin{figure}[htbp]
  \centering {    
    \includegraphics[width=\linewidth]{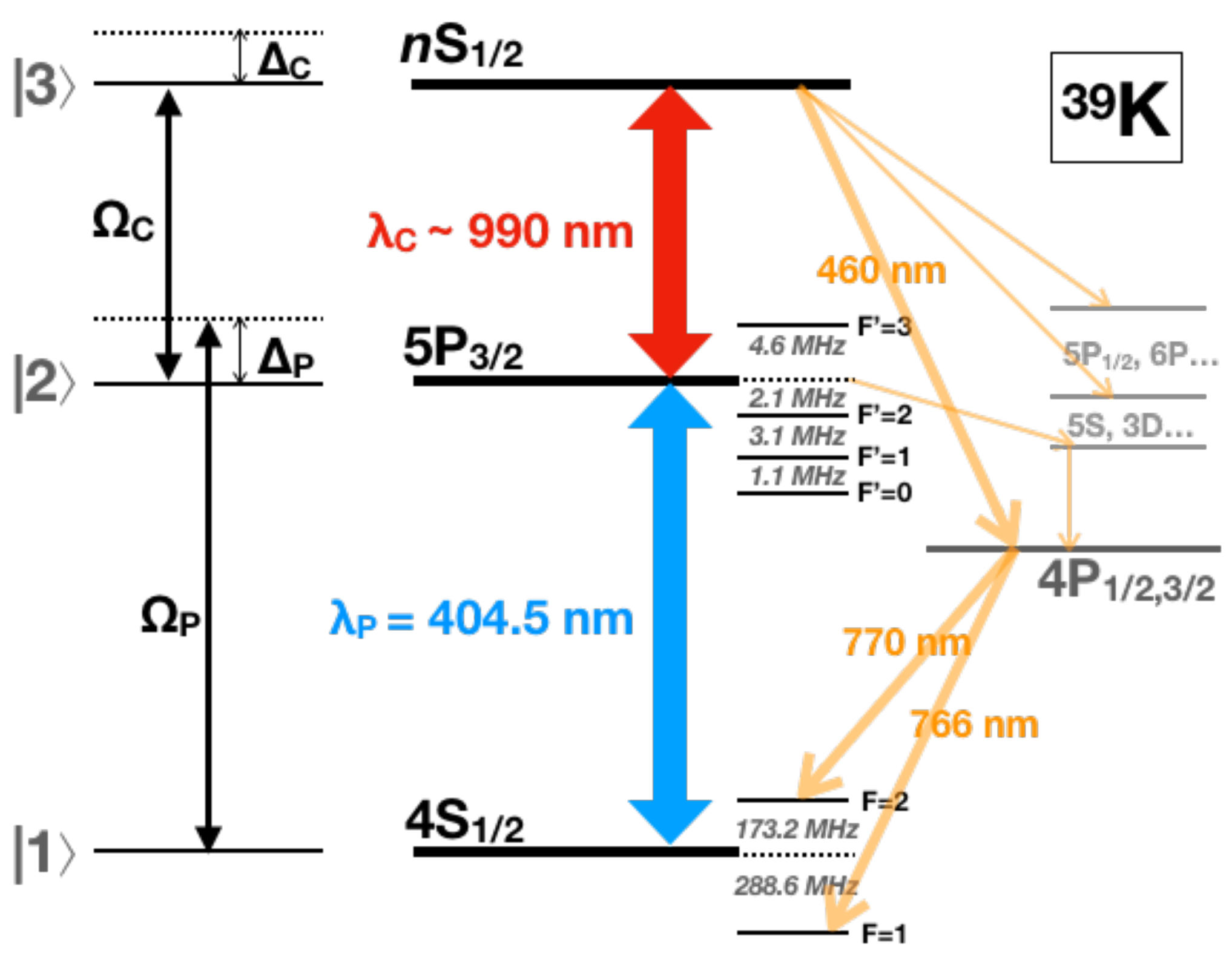}}
  \caption{Level diagram for two-photon excitation to the \emph{ns} Rydberg level using 404.5~nm and 990~nm lasers as probe and coupling laser, respectively. It shows the 4S$_{1/2}$ ground state, the intermediate $5P_{3/2}$ state, the nS$_{1/2}$ Rydberg state, and other relevant decay channels. The relevant Rabi frequencies $\Omega_p$ and $\Omega_c$ , detunings $\Delta_p$ and $\Delta_c$ are indicated.
  }
 \label{fig:level}
\end{figure}

The Rydberg excitation in a ladder-type three-level atoms using two optical fields includes two effects: two one-photon resonances (two-step) and two-photon coherent excitation. They can be considered separately in the calculations under the weak-probe approximation.
No Doppler-free feature is predicted to be observable in the inverted ladder systems \cite{Noh:2011ih,Moon:2012vl,Hayashi:2010un} under such an approximation.
In a more realistic experimental condition, the weak-probe approximation can be inadequate, thus we take all-order numerical calculation in our theoretical simulation. The density matrix is calculated numerically considering the high order effects of the coupling and probe laser strengths without any approximation.
Although the actual states involved in our experiment are more than three, as shown in Fig.~\ref{fig:level}, we simplify it using an equivalent three-level system to obtain an effective steady state solution. All the additional decay channels, such as 4P state that has a fast decay rate to the 4S ground state, are treated as part of the 3$\rightarrow$1 relaxation channel between the Rydberg state and the ground state. In comparison with the experimental results, such an equivalent three-level system is found to be reliable approach for describing the complicated systems without losing any essential charateristics.

The simplified ladder configuration used in our simulation is shown in the left of Fig.\ref{fig:level}. The optical Bloch equations are 
\begin{equation}
\begin{split}
\dot{\rho_{11}} &=i\frac{\Omega_{p}}{2}(\tilde{\rho_{21}}-\tilde{\rho_{12}})+\Gamma_{21}\rho_{22}+\Gamma_{31}\rho_{33}\\
\dot{\rho_{22}} &=i\frac{\Omega_{p}}{2}(\tilde{\rho_{12}}-\tilde{\rho_{21}})+i\frac{\Omega_{c}}{2}(\tilde{\rho_{32}}-\tilde{\rho_{23}})+\Gamma_{32}\rho_{33}-\Gamma_{21}\rho_{22}\\
\dot{\rho_{33}} &=i\frac{\Omega_{c}}{2}(\tilde{\rho_{23}}-\tilde{\rho_{32}})-(\Gamma_{32}+\Gamma_{31})\rho_{33}\\
\dot{\tilde{\rho_{12}}} &=-i(\Delta_{p}-i\gamma_{21})\tilde{\rho_{12}}+i\Omega_{p}(\rho_{22}-\rho_{11})-i\Omega_{c}\tilde{\rho_{13}}\\
\dot{\tilde{\rho_{23}}} &=-i(\Delta_{c}-i\gamma_{32})\tilde{\rho_{23}}+i\Omega_{c}(\rho_{33}-\rho_{22})+i\Omega_{p}\tilde{\rho_{13}}\\
\dot{\tilde{\rho_{13}}} &=-i(\Delta_{p}+\Delta_{c}-i\gamma_{31})\tilde{\rho_{13}}+i\Omega_{p}\tilde{\rho_{23}}-i\Omega_{c}\tilde{\rho_{12}}.\\
\end{split}
\label{eq:density}
\end{equation}

\begin{figure*}[bth]
  \centering { 
    \includegraphics[width=0.9\linewidth]{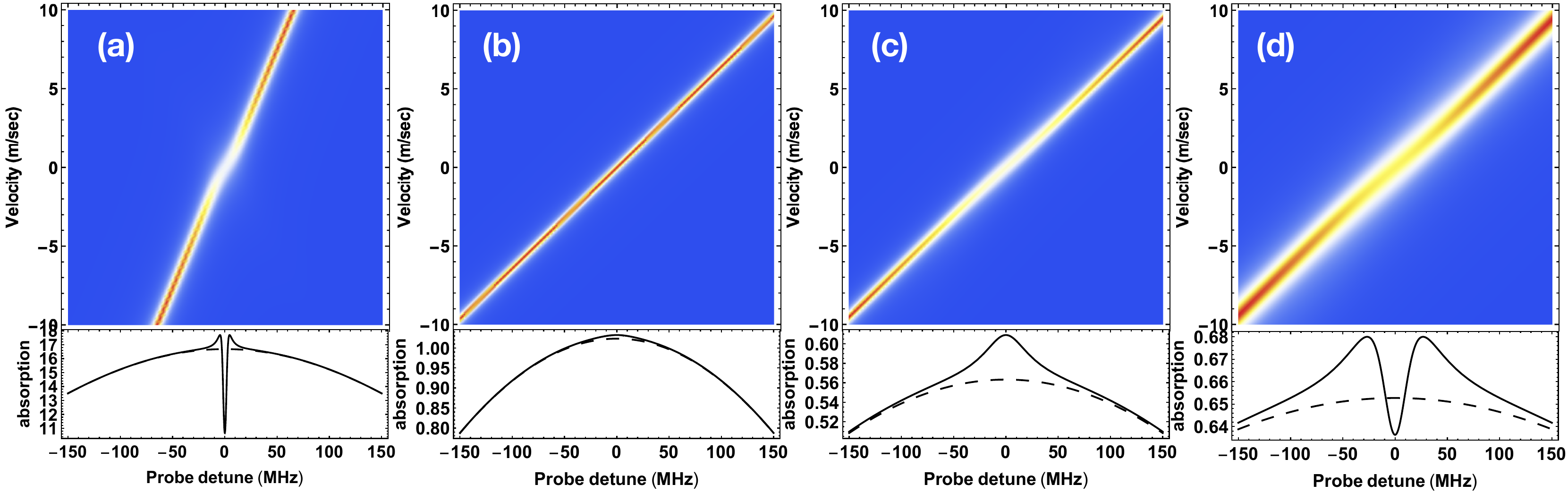} }
    \caption{The simulations of the absorption signal, where the imaginary part of $\rho_{12}$ is plotted, present the difference between non-inverted (a) ($\lambda_c$=404.5~nm and $\lambda_p$=980~nm) and inverted (b)-(d) ($\lambda_c$=980~nm and $\lambda_p$=440.5~nm) systems.
    In the upper panel illustrates velocity maps of the absorption signal. After integrating over the Gaussian-weighted signals for each atomic velocity class, the absorption signal versus the probe laser detune are shown in the solid curves (with the coupling beam) and the dashed curves (without the coupling beam).
    As a comparison, the non-inverted case (a) uses the same simulation parameters ($\Omega_p$=0.2~MHz, $\Omega_c$=12~MHz) as the inverted case (b). The simulation parameters in (c) is $\Omega_p$=1~MHz and $\Omega_c$=12~MHz. The simulation parameters in (d) is $\Omega_p$=8~MHz and $\Omega_c$=12~MHz.}
\label{fig:velocity-map} 
\end{figure*}
The levels $\Ket{1}$, $\Ket{2}$, and $\Ket{3}$, are corresponding to the 4S$_{1/2}$, 5P$_{3/2}$, and nS$_{1/2}$ states. The weak probe field and the strong coupling field are tuned close to the lower $\Ket{1}$-$\Ket{2}$ transition and the upper $\Ket{2}$-$\Ket{3}$ transition, respectively. The respective fields are denoted by: wavelength $\lambda$, Rabi frequency $\Omega$, and detuning $\Delta$. 
The spontaneous decay rate of the intermediate state $5P_{3/2}$ ($\Gamma_{21}$) is around 7.35~MHz, which is mainly attributed to three decay channels via 5S$_{1/2}$, 3D$_{3/2}$ and 4S$_{1/2}$. The spontaneous decay rate $\Gamma_{32}$ is calculated to be 1.7~kHz.

While the decay of the Rydberg state back to the ground state is primarily through the cascade $n$S$\rightarrow$4P$\rightarrow$4S cascade transition, the spontaneous decay rate $\Gamma_{31}$ is calculated to be 0.028~MHz. In our effective three-level model, $\Gamma_{31}$ also serves as an effective parameter including all the relaxation channels from the Rydberg state, such as collision quench and transit effects.
It was manually adjusted for a best fit with the experimental results, and found to be $\sim$30~MHz that is much larger than the calculated rate from the radioactive relaxation. The dephasing rates, $\gamma_{nm}$, are taken to be $\Gamma_{nm}$/2 supposing the laser linewidth $\delta$$\omega$$\sim$0. However, $\gamma_{32}$ that is related to the open decay channel and dominates the linewidth of the EIA feature was also adjusted for a best fit with the experimental results.
The parameters used in our simulation are typically: $\Gamma_{21}$=7~MHz, $\gamma_{21}$=3.5~MHz, $\Gamma_{31}$=0.16~MHz, $\gamma_{31}$=30~MHz, $\Gamma_{32}$=0.0017~MHz, $\gamma_{23}$=3.5~MHz.

In a hot cell with a Doppler-broadened medium, the probe transmission signal must be integrated over Maxwell-Boltzmann velocity distribution.
For an atom moving with a velocity \textit{v} in the same direction as the probe beam, the probe laser frequency in the moving frame of the atom is blue-detuned by ($\Delta_p$+\textit{v}/$\lambda_p$) and the coupling laser frequency is red-detuned by  ($\Delta_c$-\textit{v}/$\lambda_c$).
Hence,
\begin{equation}\label{Ndoppler}
\begin{split}
\langle\tilde{\rho_{ij}}(\Delta_{p},\Delta_{c})\rangle=\int_{-\infty}^{\infty}\tilde{\rho_{ij}}(\Delta_{p}+\frac{\omega_{p}v}{c},\Delta_{c}-\frac{\omega_{c}v}{c})N(v)dv,
\end{split}
\end{equation}
where
\begin{equation}
\begin{split}
N(v)=\sqrt{\frac{m}{2\pi kT}}e^{\frac{-mv^{2}}{2kT}},
\end{split}
\end{equation}
In Fig.~\ref{fig:velocity-map}, it evidently describes how the atoms with different velocities contribute the sub-Doppler feature to both non-inverted and inverted ladder-type cascade excitation with the wavelength mismatching. The upper parts are the density plots with atomic velocity and probe beam detuning as x and y axis. The lower parts are the integrals of all the velocity groups and the simulated observable signals. The solid lines are the absorption of the probe beam with the coupling beam, and the dashed lines are without the coupling beam. Our experiment is to utilize the modulation transfer technique (see \ref{setup}) to measure the differential signal between the ON/OFF of the coupling beam, and to remove the large Doppler background in the probe signal. Thus, the Doppler backgrounds have been subtracted from all the simulation presented in the following discussion. 

As shown in the non-inverted case in Fig.~\ref{fig:velocity-map}(a), a narrow sub-Doppler window can clearly emerge under all kinds of the conditions of probe beam power, even with a weak probe beam. However, in the inverted ladder system as shown in Fig.~\ref{fig:velocity-map}(b), which is with the same parameters, exchanging the wavelengths of probe laser and coupling laser, no sub-Doppler feature can be observed anymore, even with a strong coupling beam ($\Omega_p$=0.2~MHz and $\Omega_c$=12~MHz). It agrees with the prediction under the weak-probe approximation. In Fig.~\ref{fig:velocity-map}(c), with a stronger probe beam ($\Omega_p$=1~MHz), a relative wide sub-Doppler feature appears as a narrow absorption peak. In Fig.~\ref{fig:velocity-map}(d), while the probe beam are even stronger ($\Omega_p$=8~MHz), an EIT feature emerges to result in a complicated profile as interference between transparency and absorption. It should be noticed that the sub-Doppler width in the inverted systems is much wider than that of the non-inverted systems. The linewidth of the absorption part can be estimated by $\Delta\omega(\lambda_c/\lambda_p)$, where  $\Delta\omega$ is the linewidth of one-photon transition. The $5P_{3/2}$ hyperfine splitting is 10.8~MHz. Because of the coupling field, the power broadening is particularly significant in our experiment. Taking $\Omega_c$=12~MHz into account, $\Delta\omega\sim$23~MHz and the absorption linewidth is estimated to be $\sim$55~MHz that is in good agreement with the observed linewidth. On the other hand, there is no simple method enable a quick estimation of the width of transparent window that involved coherent interference phenomenon. 

In both of the non-inverted and inverted schemes, the sub-Doppler window appears in the situation, where the two-step condition, i.e.$\Delta_c$+$\Delta_p$=($\omega_p$-$\omega_c$)v/c, is satisfied. Yet the size and the width of the transparency window is severely influenced by the wavelength ratio \cite{Shepherd:1996uy}. Based on our simulation, the inverted ladder types are certainly at unfavorable situations for narrowing the coherence window. That explains why the sub-Doppler features are expected to be vanished under typical weak-probe approximations. However, beyond such an approximation and with all-order numerical calculations, a sub-Doppler feature has been predicted to be feasible by \cite{GEABANACLOCHE:1995wt}. The sub-Doppler feature can be a transparency window, an absorption dip, or a combination of both, as experimentally demonstrated by \cite{Urvoy:2013if,Xu:2016df}.

\section{Experimental Setup}
\label{setup}

\begin{figure}[htbp]
  \centering { 
\includegraphics[width=\linewidth]{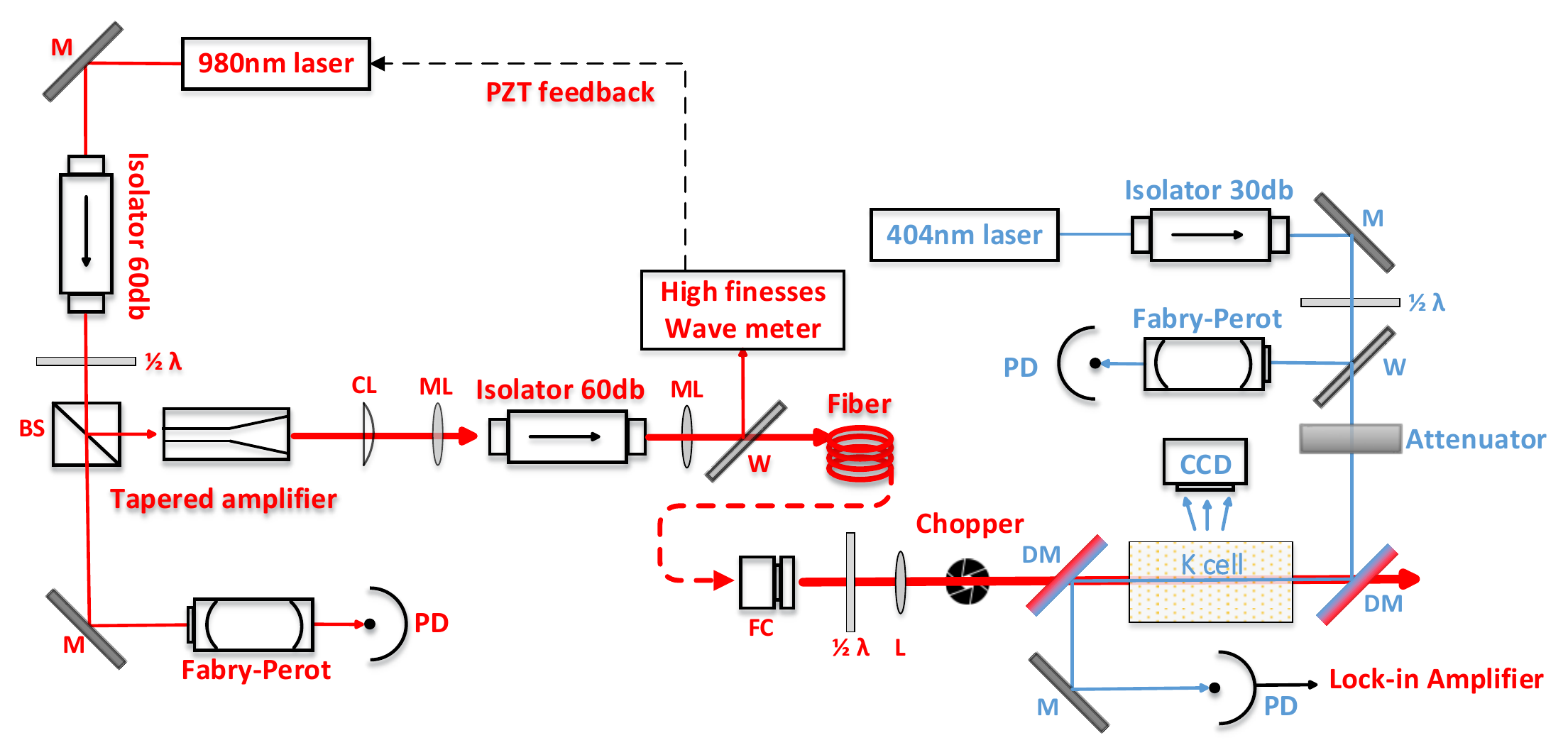}} 
  \caption{Experimental setup for Rydberg optical excitation in a hot cell. PD: photo detector, M: mirror, W: window, DM: dichroic mirror, L: f=40 cm lens, BS: bean splitter, ML: mode matching lens, CL: cylindrical lens, FC: fiber collimator.}
   \label{fig:setup} 
\end{figure}

The experimental setup shown schematically in Fig.~\ref{fig:setup} is composed of two tunable laser systems and a hot cell Rydberg spectrometer. The ultra-violate light at 405~nm, as a probe beam, is to excite $4S_{1/2}\rightarrow 5P_{3/2}$ transition, while the near-infrared laser, as the coupling beam, with a tuning range from 970~nm to 990~nm is to excite $5P_{3/2}\rightarrow nS_{1/2}(nD_{3/2,5/2})$ transition. Both probe and coupling lasers are generated by home-made external cavity diode lasers (ECDL) with Littrow configuration. The lasers are capable of mode hop free range over 5~GHz with the current feed-forward technique.
In order to further amplify the coupling laser power to $\sim$1~W, the near-infrared laser output from ECDL is injected into a GaAs based tapered amplifier chip (Coherent, TA-0976-2000) as a Master Oscillator Power Amplifier (MOPA) configuration. Then, the coupling beam passes through a polarization-maintaining optical fiber and is focused in the center of the cell with a beam diameter of 1.6~mm spatially overlapped with the probe beam with a beam diameter of 0.8~mm. This configuration ensures that the entire probe beam is covered by the coupling light. The maximum powers reaching to the interacting region of the cell are typically 25~mW for the probe ($\Omega_p$=21.2~MHz) and 350~mW for the coupling ($\Omega_c$=13.2~MHz). The coupling and probe beams are arranged in counter-propagation. The probe and coupling beams are linearly polarized in parallel to each other.

The vapor cell (Thorlabs, GC25075-K) is 7.18~cm in length, containing 93.3\% $^{39}$K and 6.7\% $^{41}$K, and its temperature is stabilized at 125$^{\circ}$C with a fluctuation less than 1~$^{\circ}$C. The probe and coupling beams are separated after traveling through the vapor cell by two dichroic mirrors. The probe beam is detected using a photodiode to perform absorption spectroscopy. Meanwhile, the fluorescence signal is simultaneously monitored using a CCD camera with a narrow bandpass interference filter of 405~nm, 766~nm, or 460~nm. Our spectrometer utilizes modulation transfer techniques to remove Doppler-broadened profile and to improve the signal-to-noise ratio. The coupling laser is amplitude modulated by a mechanical chopper at 2-3~kHz, and the probe transmission is demodulated using a lock-in amplifier.

The coupling laser is locked to different frequencies for excitation to the Rydberg states (n=27-70) using software feedback control by cooperating with a high precision wavemeter (HighFinesse, WS6), which is capable of measuring the absolute frequency with an uncertainty less than 30~MHz and with a relative stability better than 1~MHz. The frequency of the probe laser is measured using either a second HighFinesse wavemeter or another wavemeter with a 1~GHz accuracy (Bristal 521).

\section{Rydberg Excitation in a Hot Cell}

\begin{figure}[htp]
    \includegraphics[width=\linewidth]{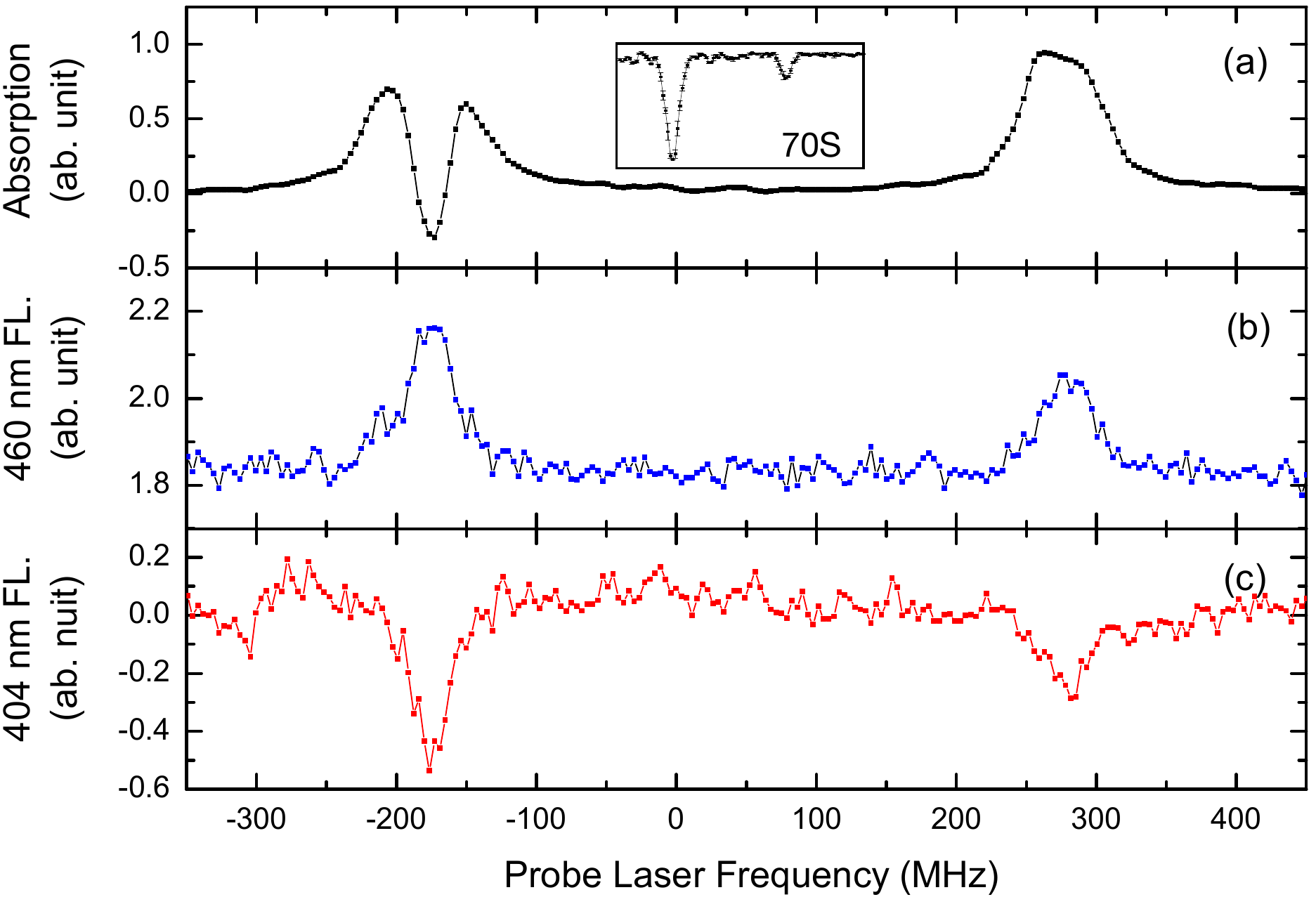}
  \caption{The two-photon excitation spectrum observed by probe laser absorption (a), 405~nm fluorescence (b) and 460~nm fluorescence (c). The coupling laser frequency are resonant to $\rm{5P\rightarrow26S}$. The two peaks are corresponding to $\rm{4S(F=2)\rightarrow5P\rightarrow26S}$ (left) and $\rm{4S(F=1)\rightarrow5P\rightarrow26S}$ (right). The inset is the absorption spectrum of 70S Rydberg state and shows clear dips on both F=1 and 2 manifolds} 
  \label{fig:spectrum}
\end{figure}

A typical experimental probe transmission spectrum is shown in the top of Fig.~\ref{fig:spectrum} with a 0.4~mW probe laser and a 300~mW coupling laser when the probe frequency scans over 4S-5P transition resonance. The fluorescence spectra of 404~nm and 460~nm are recorded simultaneously to help identify the population decaying channels. In the probe transmission spectrum, the Doppler background absorption is removed using modulation transfer technique. The left and right spectral signal are the excitation from F=2 and F=1 hyperfine structure of the 4S state, respectively. Both spectral signals include enhanced absorption part, but the transparent window is less pronouncing for F=1, because the Rabi frequency for F=1 is smaller than that for F=2 using the same probe power. The transparent window becomes observable in a higher Rydberg state, as shown in the inset of Fig.~\ref{fig:spectrum} and Fig.~\ref{fig:var-state-simu}. Despite the spectral features of these two transitions manifest differently, they can be simulated using the same rate mechanism. 

\begin{figure}[hbtp]
  \centering { 
    \includegraphics[width=80mm]{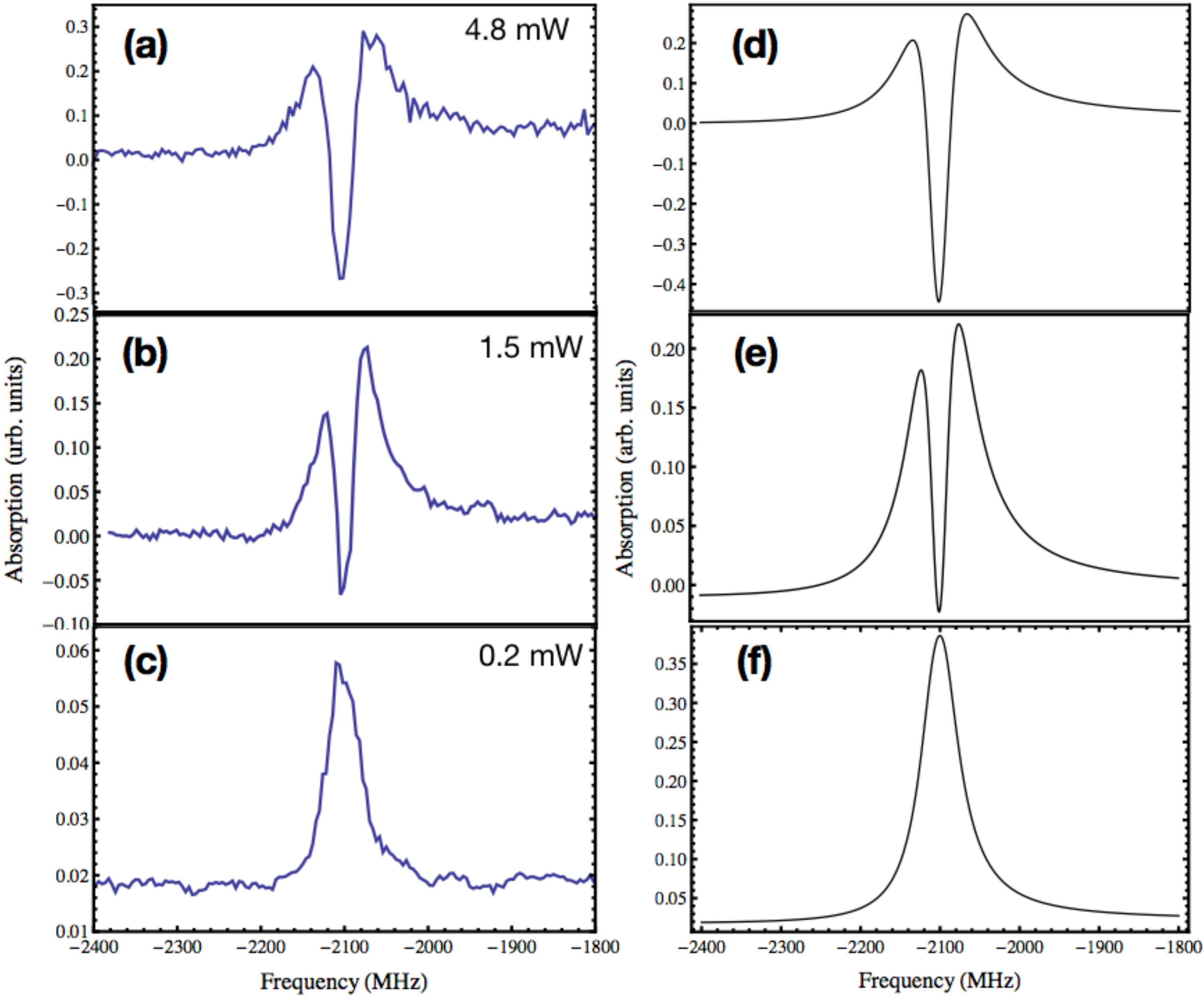}}
  \caption{(a)-(c) are the experimental results of the probe laser absorption spectra for $\rm4S_{1/2}(F=2)\rightarrow5P_{3/2}\rightarrow25S_{1/2}$ using various probe power strengths (4.8~mW, 1.5mW, and 0.2mW), while the coupling laser Rabi frequency is fixed at 12~MHz. (d)-(e) are the simulations ($\rm Im[\rho_{12}]$) corresponding to (a)-(c)}
  \label{fig:sim-exp}
\end{figure}

Figure~\ref{fig:sim-exp} illustrates the simulated sub-Doppler features using the parameters mentioned Section \ref
{set:model}. A good agreement between our experimental observations and numerical simulation was presented. It is initially an absorption peak at low probe power region, then transformed to comprising a transparency window at high probe power region. The width of the transparent window is measured to be several tens of MHz.

\begin{figure}[htp] 
\includegraphics[width=\linewidth]{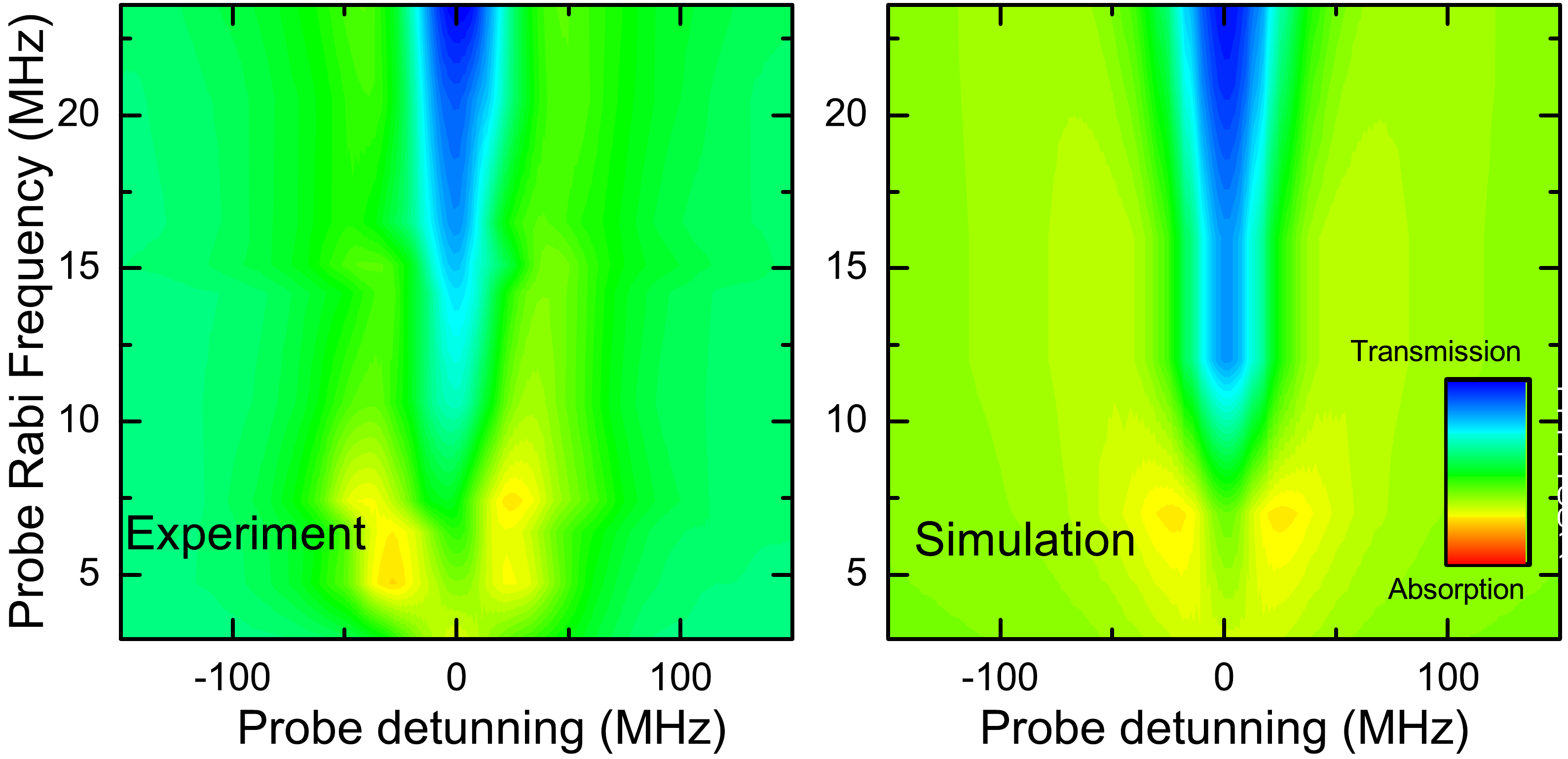}
  \caption{The probe power dependence of EIT signal ($\rm4S_{1/2}(F=2)\rightarrow25S$) for experimental results and simulations. The probe laser Rabi frequency $\Omega_p$ varies from 3 to 22~MHz and the coupling Rabi frequency is fixed at $\Omega_c$=13~MHz with the relaxation parameters $\Gamma_{21}$=7, $\gamma_{21}$=3.5, $\Gamma_{31}$=0.16, $\gamma_{31}$=30, $\Gamma_{32}$=0.0017, $\gamma_{32}$=3.5~MHz.}
\label{fig:var-probe-simu}
\end{figure}

\begin{figure}[htp] 
\includegraphics[width=\linewidth]{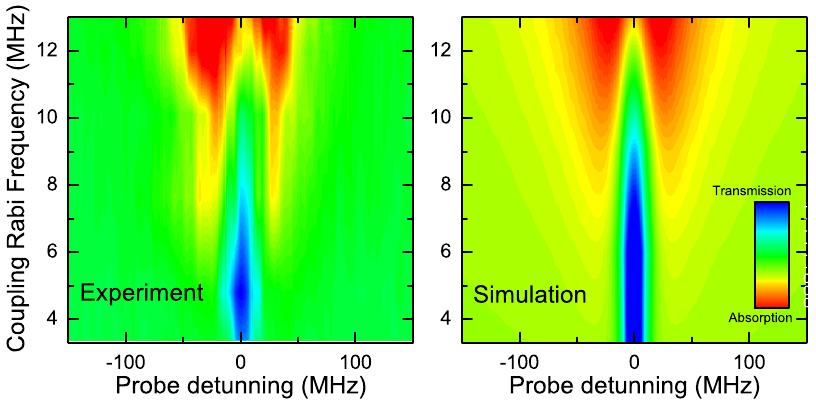}
  \caption{The coupling power dependence of EIT signal ($\rm4S_{1/2}(F=2)\rightarrow25S$) for experimental results and simulations. The coupling laser Rabi frequency $\Omega_c$ varies from 3 to 13~MHz and the probe Rabi frequency is fixed at $\Omega_p$=7~MHz (2~mW) with the exactly same relaxation parameters used in Fig~\ref{fig:var-probe-simu} }
\label{fig:var-coupling-simu}
\end{figure}

\begin{figure}[htp]

\includegraphics[width=\linewidth]{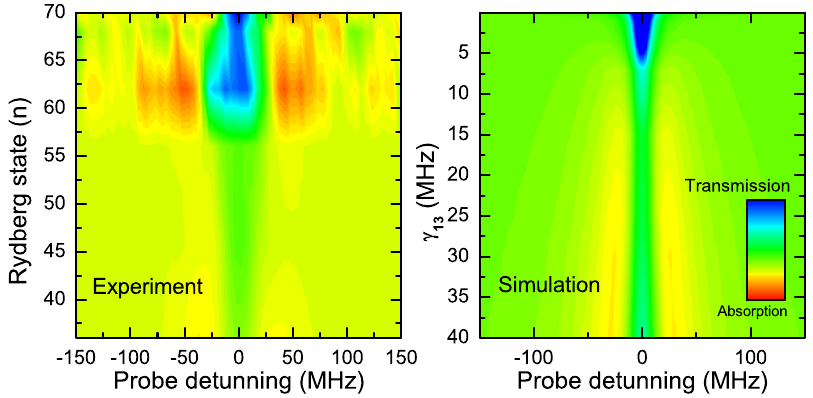}
  \caption{The density plots of EIT signal for various $n$ Rydberg states in comparison with that of the various dephasing rate. The experimental data are taken for the Rydberg state with n=36, 46, 56, 62, 68, 70. The all-order numerical simulations is for the EIT signal with a $\gamma_{13}$ in the range of 1-40~MHz and with $\Gamma_{21}$=7~MHz, $\gamma_{12}$=3.5~MHz, $\Gamma_{31}$=0.16~MHz, $\Gamma_{32}$=0.0017~MHz, $\gamma_{23}$=3.5~MHz, $\Omega_p$=6.4~MHz and $\Omega_c$=10.2~MHz.}
\label{fig:var-state-simu}
\end{figure}

For a specific transition, the competition between the strength of the transmission and absorption parts depends on Rabi frequencies of the probe and the coupling lasers, as shown in Fig.~\ref{fig:var-probe-simu} and Fig.~\ref{fig:var-coupling-simu}, respectively. We obtain a good agreement between our all-order numerical simulation and experiments for both the cases. Figure~\ref{fig:var-probe-simu} represents that interference feature is changed from an absorptive doublet to a transparent window at higher probe Rabi frequency. It can be understood as the power broadening resulting in overlap and interference between the Autler-Townes doublet.
In Fig.~\ref{fig:var-coupling-simu}, the transmission dip clearly enhanced in the situation where the coupling Rabi frequency is smaller than 8~MHz. As the coupling Rabi frequency is increased, the absorption components grow up. It can be attributed to the transformation from EIT to Autler-Townes splitting due to the strong coupling intensity, thus the quantum interference is smeared out.

The observed dephasing rate $\gamma_{31}$, a few tens~MHz, is much larger than 12~kHz, which is directly deduced from the intrinsic lifetime of the states. Such a fast dephasing rate has also been observed and discussed in other experiments \cite{Raitzsch:2009ep, Zhang:2018dg}. This broadening might be caused by the transit effect, the optical pumping, the velocity-change collision in the hot cell, and the population trapping in the ground hyperfine states. In a closed system, optical pumping will depletes the ground-state atoms in less than $\mu$s. However, in a thermal ensemble that is an open system, the atoms move into and out of the spatial and momentum space, where atoms interact with lasers. The ground state population reaches to a steady state that is a balance of this incoherent dynamic process. Thus, a large dephasing rate was introduced to allow us to employ a steady state simulator for a dynamic open system.

The result shows our simple model cooperating very well with the experimental results. In this model, the hyperfine manifold of $5P_{2/3}$ is approximated as a single intermediate level. As the observed signal is a sum of all possible excitation routes, the approximation is then eligible for the intermediate manifold with small energy splittings and a relatively wide linewidth.

 It should be noted that the Fig.8 is plotted using discrete principle number against probe detuning. Even so, it still shows a continuous transformation in the spectrum from low $n$ to high $n$, where both of the EIT transmission part and absorption part is attenuated in the lower Rydberg states. There are several factors responsible for the spectrum transformation between different Rydberg states. Besides excitation rate, one of the dominant factors is dephasing rate, which is caused by the presence of interactions between Rydberg atoms. As a consequence, the dephasing rate ranged from 1-40 MHz is plotted to compare with the experimental data. Their trends in EIT transmission part can be found to the same. The similar phenomenon was also observed in \cite{han2016spectral}.

\section{Level energy measurement}
\begin{figure}[htp] 
\includegraphics[width=0.8\linewidth]{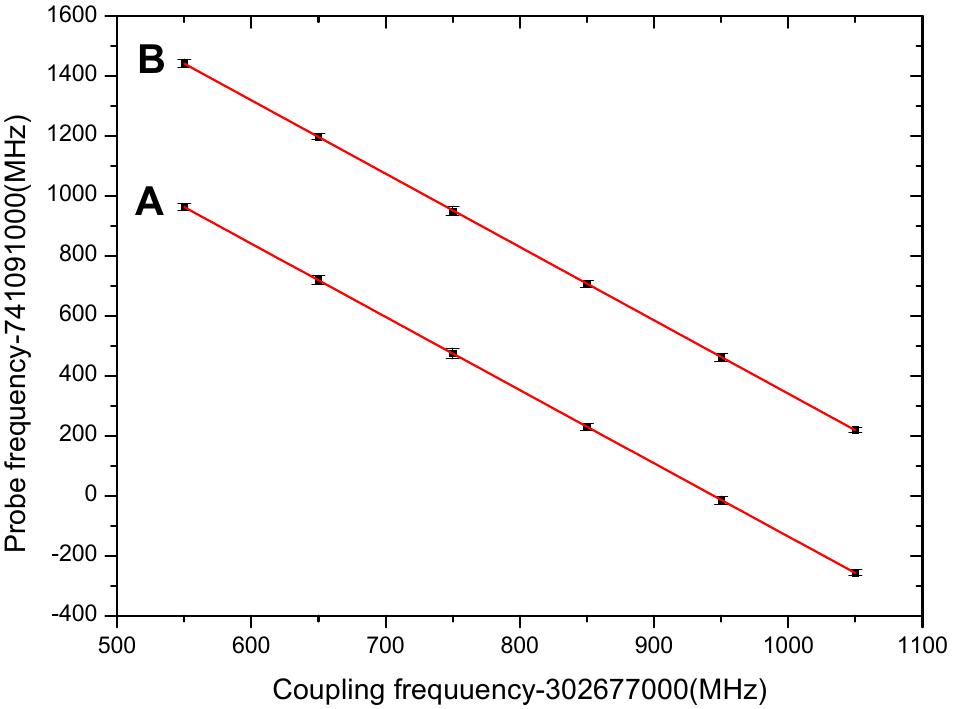}
  \caption{The frequency dependence between the probe frequency detune and coupling frequency detune, where A is for 4S$_{1/2}$(F=2)-26S and B is for 4$S_{1/2}$(F=1)-26S, along with linear fitting curve (red line).} 
  \label{fig:freq-slope}
\end{figure}

There are two primary excitation paths happening in a three-level cascade system: two-photon and two-step excitations. They can be distinguished by the frequency dependence between the coupling and probe laser frequency detunings. The photon energies in two-photon excitation need to satisfy the following conditions:
\begin{equation}\label{eq:two-photon}
\begin{split}
&f_c+f_p=f_{12}+f_{23},\\
&\Delta_p=-\Delta_c.
\end{split}
\end{equation}
On the other hand, the condition for two-step excitation, where two one-photon excitations are included, should satisfy the following conditions:
\begin{equation}\label{eq:two-step}
\begin{split}
&f_p=f_{12}(1-\frac{v}{c})~\textrm{and}~f_c=f_{23}(1+\frac{v}{c})\\
&\Delta_p=-\frac{f_{12}}{f_{23}}\Delta_c,
\end{split}
\end{equation}  
where $\Delta_p$ (=$f_{12}-f_p$) and $\Delta_c$ ($=f_{23}-f_c$) are frequency detunings for the probe and coupling, respectively. In the case of two-step excitation, the  co-propagating configuration of probe and coupling beam could also provide a sub-Doppler feature, as \cite{Urvoy:2013if}.
To clarify the dominating path which results in the sub-Doppler feature in our observation, a series of the absorption spectrum of the Rydberg state n=26 with various coupling laser frequencies were taken. As shown in Fig.\ref{fig:freq-slope}, the frequency shift of the resonant probe laser is linearly dependent on the frequency detuning of the coupling laser with a slope of -2.448 ($=-f_{12}/f_{23}$). It evidences that the sub-Doppler feature arises primarily due to the velocity group of atoms that are simultaneously resonant to both coupling and probe lasers, that is, the two-step excitation using 
\begin{equation}\label{eq:shift}
\frac{v}{c}=\frac{\Delta_p}{f_{12}}=-\frac{\Delta_c}{f_{23}}
\end{equation}
,and the frequency of the Rydberg state is then derived as:
\begin{equation}\label{eq:shift}
f_{Rydberg}=f_{12}+f_{23}=f_{12}+f_c-\frac{f_c}{f_{12}}(f_{12}-f_p)
\end{equation}
In combination with the precisely measured 4S-5P transition frequency $f_{12}$ \cite{Sansonetti:2008ej}, the sub-Doppler signal with a good signal-to-noise ratio allows precision measurements for the Rydberg states up to 70S (Table.\ref{tab:freq}) by simultaneously measuring $f_p$ and $f_c$. In our experiment, both lasers are either locked to the wavemeter or measured in real time using a wavemeter, since all of the uncertainties are limited by the accuracy of wavelength meter. While the frequency uncertainties of both $f_p$ and $f_c$ are 30~MHz using the Highfinesse wavemeter and $f_c/f_{12}=1/2.45$, the uncertainty of $f_{Rydberg}$ is 32~MHz for n=20-26. For the others, the less accurate results are due to the less accurate $f_p$ using a wavemeter with an accuracy of only 1~GHz. The unresolved nS (F=1 and F=2) Rydberg hyperfine may result in a broadened symmetrical line shape that could introduce additional uncertainty to the measurement. However, it is too small to be taken into account in our experiment. Our results are in good agreement with previous two-photon spectroscopy. With several newly measured energy levels of the high $n$ ($>$55) states, our experimental results enable to derive the ionization energy ($E_i$). With sufficiently high principal quantum numbers, the energies of Rydberg level are given by: 
\begin{equation}
E_{n,l,j}=E_i-\frac{R_K}{[n-\delta(n,l,j)]^2},
\end{equation}
where ${R_K}$ is the Rydberg constant and only the low-order quantum defect $\delta(n,l,j)$ being taken into account under the approximation of the modified Rydberg-Ritz parameters \cite{Mack:2011ij}
\begin{equation}
\delta(n,l,j)=\delta_0+\frac{\delta_2}{(n-\delta_0)^2}.
\end{equation}
The derived ionization energy of potassium is $ E_i=35009.87(6)~\rm{cm}^{-1}$, which is in agreement with the previous reported values \cite{Lorenzen:1981wl, Thompson:1983dy}.

\begin{table}[htb]
\begin{center}
\caption {The energy level measurements of Rydberg states from 20S to 70S excited from the ground state 4S$_{1/2}$F=2. The frequency uncertainties are limited by the accuracy of wavelength meter. While the results for 20S-26S are measured with two high precision wavelength meters (with an accuracy of 30~MHz or $\rm0.001~cm^{-1}$), others are measured with a normal wavelength meter (with an accuracy of 1~GHz or $\rm0.03~cm^{-1}$). All units are in cm$^{-1}$.} 
\begin{tabular}{llcc}
\hline
\hline
level & This work & \cite{Thompson:1983dy}& \cite{Xu:2016df}\\ 
\hline
20S        & 34664.2114(11)  & 34664.2127(30) &   \\ 
21S        & 34699.9642(11)  & 34699.9574(30) &  \\ 
22S        & 34730.4395(11)  & 34730.4327(30) &  \\ 
24S        & 34779.3102(11)  & 34779.3240(30) &   \\ 
25S        & 34799.0695(11)  & 34799.0620(30) &   \\ 
26S        & 34816.3926(11)  & 34816.3824(30) &  34816.4059(27)  \\
27S        & 34831.66(3)    & 34831.6574(30) &   34831.6769(27)\\
28S        & 34845.17(3)    & 34845.1864(30) & 34845.1755(27)  \\
36S        & 34913.85(3)    & 34913.8546(30) &    		  \\ 
46S        & 34952.64(3)    & 34952.6446(30) &   \\ 
50S        & 34961.78(3)    & 34961.8083(30) &   \\ 
54S        & 34968.92(3)    &&   \\ 
56S        & 34971.89(3)    &&   \\
62S        & 34979.13(3)    &&   \\
66S        & 34982.83(3)    &&   \\
68S        & 34984.36(3)    &&   \\
70S        & 34985.93(3)    &&   \\ 
\hline
\hline
\label{tab:freq}
\end{tabular}
\end{center}
\end{table}

\section{Rydberg Excitation in Cold Atoms}
As the discussion and the experimental results shown above, it can be concluded that the sub-Doppler EIT signal appearing in the inverted ladder type scheme relies on the two-step excitation by the Doppler shift mechanism of the thermal atoms. With cold atomic ensemble, which enables only two-photon excitation by suppressing the two-step excitation because of its narrow velocity distribution.

It had been found in many experiments using cold atomic ensembles that the atom number of the trap decreases unexpectedly large with Rydberg state excitation. They might be originated from collision with ionized electron \cite{RobertdeSaintVincent:2013kk}, spontaneous avalanche dephasing \cite{Boulier:2017ul}, and optical pumping \cite{PhysRevA.93.043425}.
In the section, we report the Rydberg excitation trap loss spectroscopy in a potassium MOT to investigate possible mechanism of such a large loss.  
In our experiment, the cold potassium $\rm^{39}K$ atoms are trapped in a MOT with a total number of $>10^9$, corresponding to a number density of $\rm\sim10^{10}/cm^3$. The temperature of the atomic cloud is $\sim$200~$\mu$K, which results in a Doppler width of only 1.2 MHz for 405~nm. In our MOT, the 766~nm cooling laser is tuned to $4S_{1/2}(F=2)\rightarrow4P_{3/2}(F=3)$ with a 12~MHz red-detuning and the repump laser is on the resonance of $4S_{1/2}(F=1)\rightarrow4P_{3/2}(F=2)$ transition. The total intensity of all the beams is $\rm6~mW/mm^2$. The ratio of cooling and repump beam is 3:1. The gradient of the magnetic field is 10~G/cm.

\begin{figure}
\includegraphics[width=\linewidth]{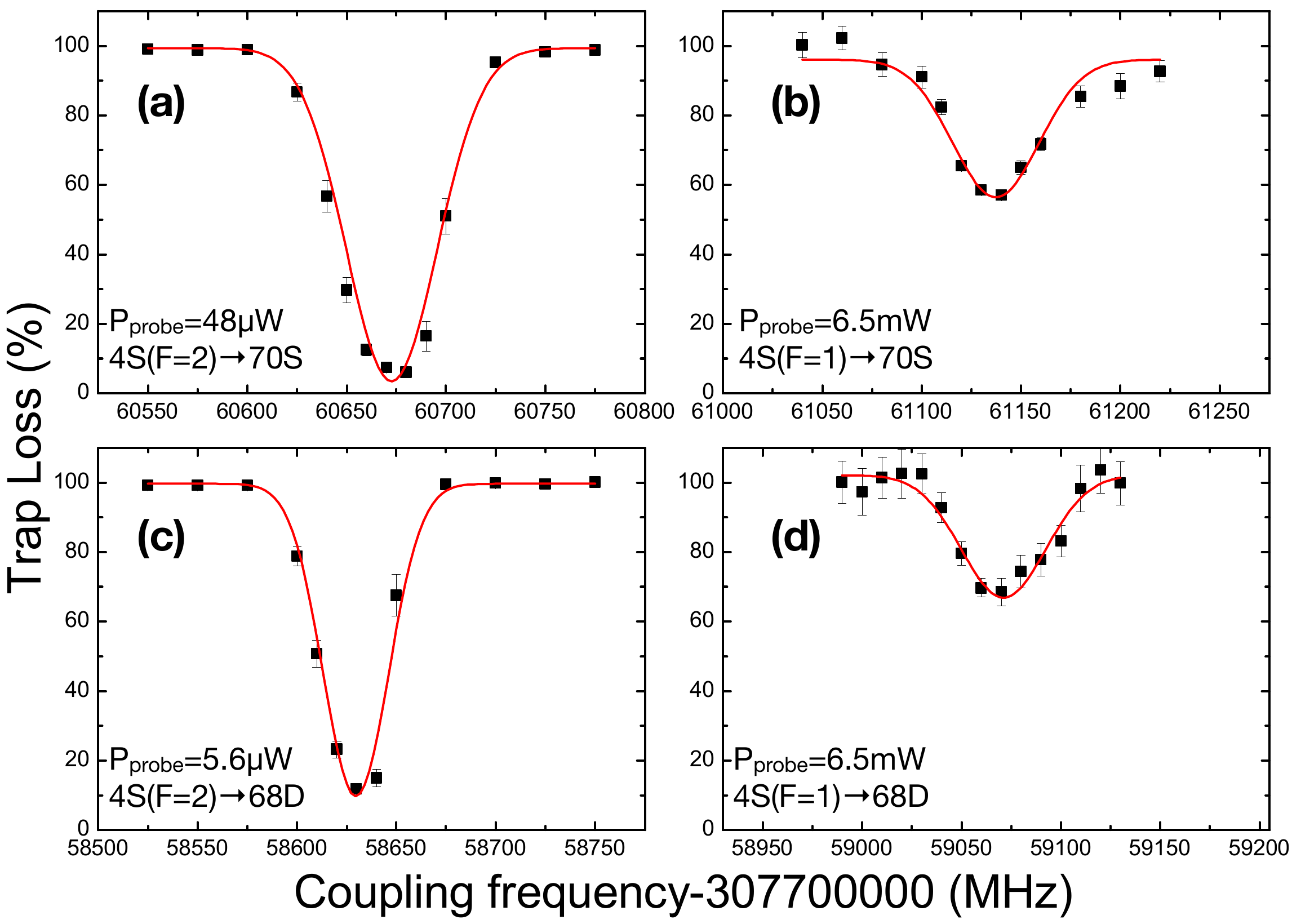}
\caption{The steady state trap loss measurement. To excite the potassium MOT atoms to 70S and 68D, the 404~nm probe laser is frequency locked to the saturation signal of $\rm 4S_{1/2}(F=2)\rightarrow5P_{3/2}$ transition in the hot cell. The $\Delta_p$ seen by $\rm 4S_{1/2}$ F=2 is 10~MHz while the $\Delta_p$ is 455~MHz seen by $\rm 4S_{1/2}$ F=1. The coupling laser power is maintained to be 365~mW for all the measurements, but the probe powers are 48~$\mu$W for (a), 6.5~mW for (b) and (d), and 5.6~$\mu$W for (c). The measurements are taken after all the lasers and MOT being turned on for 10 sec to reach the steady state. The centers are found to be at the two-photon resonances, $\Delta_p+\Delta_c=0$.}
\label{Fig:mot}
\end{figure}

\begin{figure}
\includegraphics[width=\linewidth]{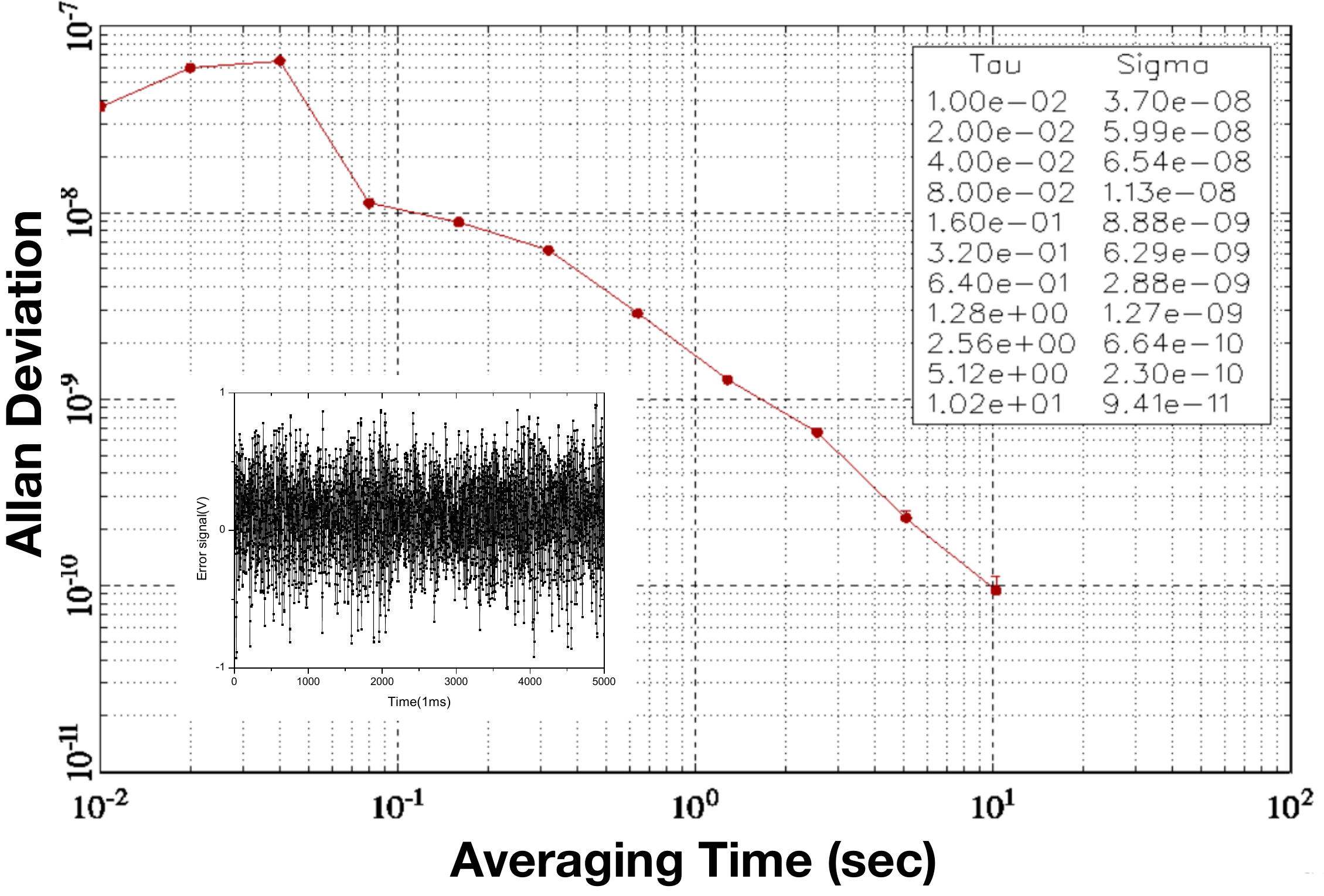}
\caption{The Allan deviation of the probe laser while it was locked to the saturation signal of the transition $4S_{1/2}(F=2)\rightarrow5P_{3/2}$.}
\label{Fig:allan}
\end{figure}

Instead of using absorption signal to observe EIT in hot cell, the steady state trap loss was measured in the cold atoms by observing the 760~nm fluorescence signal from the MOT using a CCD camera while the probe and coupling beams are simultaneously sent to the MOT with beam sizes of 2.8~mm (FWHM) in a counter-propagating arrangement. The trap loss spectra of 70S and 68D transitions are illustrated in Fig~\ref{Fig:mot}. The peak position of the trap loss satisfies the two-photon resonances condition $\Delta_p+\Delta_c=0$. The trap loss signal of each measurement was obtained under the condition that the loss-capture equilibrium was reached, while both of the coupling and the probe lasers were stabilized at specific frequencies.
The probe laser was side-locked to the saturation signal of the transition $4S_{1/2}(F=2)\rightarrow5P_{3/2}$ with a frequency deviation of 1.5~MHz at 10~ms integration time, as shown in the Allan plot, Fig.~\ref{Fig:allan}.
Hence, the probe frequency is $\sim$10~MHz ($\Delta_p$) away from the $4S_{1/2}(F=2)\rightarrow5P_{3/2}$, and equivalent to $\sim$466~MHz away from 4S$_{1/2}(F=1)\rightarrow5P_{3/2}$. The coupling frequency, measured by the wavemeter, was scanned over the frequency range of the $5P_{3/2}\rightarrow 68D$ and $70S$. 
For comparison, the trap loss was normalized to the baselines of the coupling laser far away from the resonances.
With the coupling laser with 365~mW and $\Delta_p\sim$10~MHz, we observed that only few $\mu$W of the probe laser power is high enough to cause a 100\% loss, as shown in  Fig.~\ref{Fig:mot}(a) and (c), where 48~$\mu$W and 5.6~$\mu$W was applied for 70S and 68D, respectively. In the off-resonance conditions $\Delta_p\sim$466~MHz in Fig.~\ref{Fig:mot}(b) and (d), a relatively high power of the coupling laser, 6.5~mW, was used  to induce an observable loss. And, the loss rate of both 70S and 68D are the same under such a condition.

In the case of far off resonance where the detuning $\Delta_p$ is much larger than the linewidth, Rydberg states are less efficiently populated. 
Therefore, the excitation optical field can be treated as a perturbation of the atom system, which can be most conveniently described by the dressed states \cite{PhysRevA.93.043425, Fleischhauer:2005ti}:
\begin{equation}
\ket{\hat g}=(1-\frac{\epsilon_e^2}{4}-\frac{\epsilon_r^2}{4})\ket{g}+\epsilon_e\ket{e}+\epsilon_r\ket{r},
\label{eq:dressed}
\end{equation}
where $\ket{g}$, $\ket{e}$, and $\ket{r}$ denote the undressed ground state, intermediate state, and Rydberg states, respectively.
The 4S ground state is thus linked to the Rydberg state through a two-photon process with its energy level shifted by ac-Stark effect. While the laser cooling and trapping force of the MOT relies upon 4S-4P transition, the 4S state level shift reduces the MOT capture rate. The ac-Stark shift is proportional to the effective two-level Rabi frequency ($\sim$$\Omega_p$$\Omega_c$/$\Delta_p$). It could explain why a higher probe laser power is required for the far off resonance cases (Fig~\ref{Fig:mot} (b) and (d)) to induce observable trap losses. In addition, the 70S and 68D Rydberg states suffer similar amounts of trap losses because their transition dipole moments are about the same.

In the case of on resonance where the detuning $\Delta_p$ is within the linewidth, for example in Fig~\ref{Fig:mot} (a) and (c), a significant amount of the ground-state population is transferred to the Rydberg states and a few $\mu$W probe laser is sufficient to induce a considerable trap loss. The 70S and 68D states excitation rate are calculated being comparable, however, 70S state still requires a probe power an order larger than 68D state to induce the same amount of trap loss. It is suggested that several proposed loss mechanisms, including the optical pumping effect \cite{PhysRevA.93.043425}, the collision with ionized free electron \cite{PhysRevA.89.022701,PhysRevLett.110.045004}, and the dephasing \cite{PhysRevA.96.053409}, which are related to the Rydberg excitation rate, cannot fully explain such a difference.

\begin{figure}
\includegraphics[width=\linewidth]{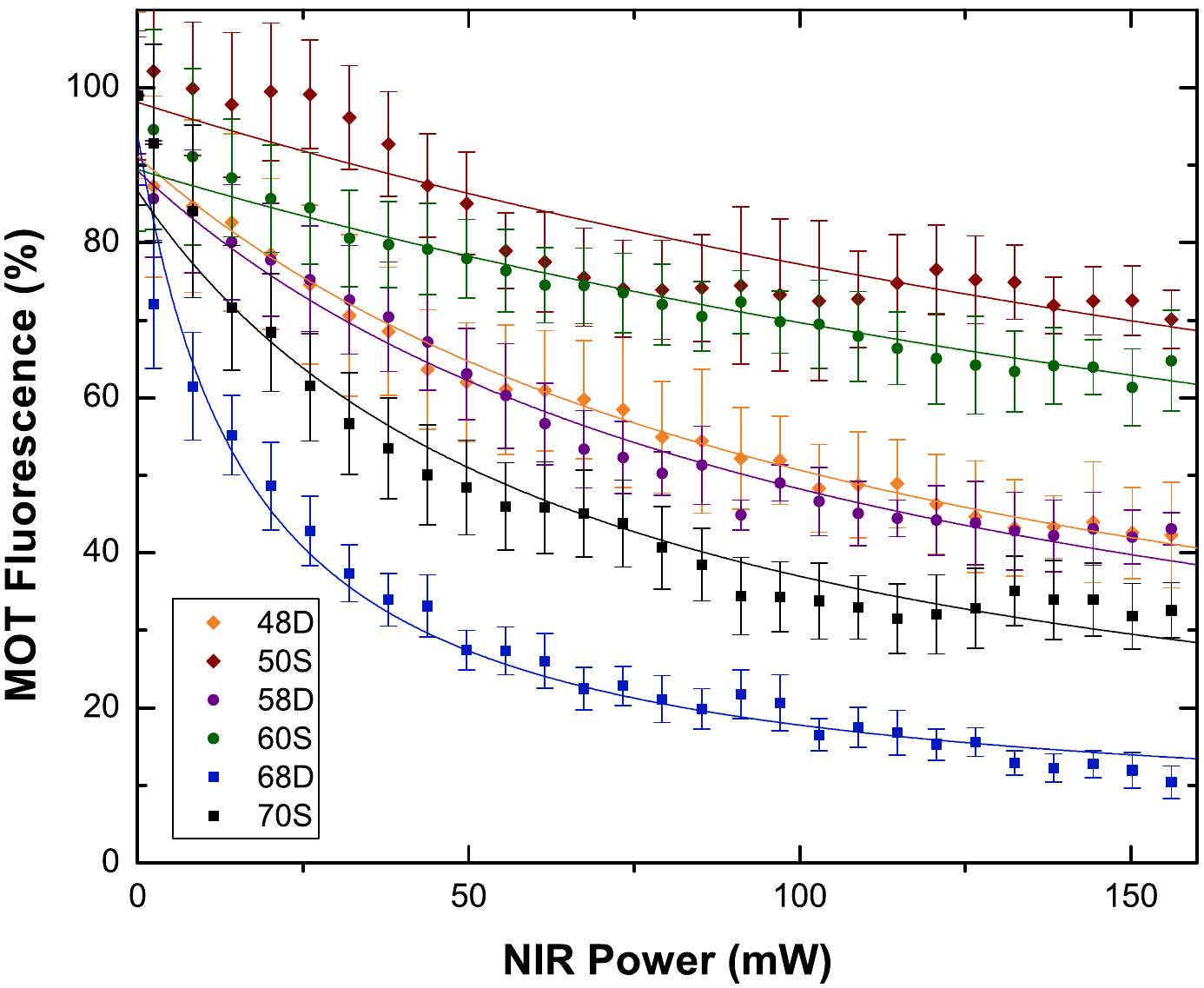}
\caption{The trap losses versus the coupling laser power for 70S, 68D, 60S, 58D, 50S, and 48D. The probe laser power was kept the same for all the measurement. While the detuning from the intermediate state 5P is 20~MHz, all data points are measured under the situation where two-photon resonance is maintained to ensure there is the maximum trap loss. The experimental data were fitted based on Eq.~\ref{eq:snloss}}
\label{Fig:mot-state}
\end{figure}

To further investigate the cases of on-resonance, the trap losses of six different Rydberg states were measured with various coupling laser powers, as shown in Fig.~\ref{Fig:mot-state}. To stress the trap loss difference affected by S and D Rydberg states, they can be divided to three pairs, i.e. 70S vs. 68D, 60S vs. 58D, and 50S vs. 48D. Each pair was chosen to be with the closest level energies, whose differences for each pair are 2~GHz, 3.3~GHz, and 5.8~GHz, respectively.
For each data point, the detuning from the intermediate state was $\Delta_p\sim$20~MHz, and the two-photon resonance condition, $\Delta_c+\Delta_p=0$, was preserved by precisely controlling the probe and coupling laser frequency. While the probe laser was locked to the Doppler-free crossover saturation signal of $\rm{4S_1/2(F=1 \& F=2)\rightarrow5P_{3/2}}(F=1,2)$, and shifted to the resonance of $\rm{4S_1/2(F=2)\rightarrow5P_{3/2}}(F=1,2)$ with 20~MHz detuning using an acousto-optic modulator (AOM). The  coupling laser was locked to the wavemeter to match the two-photon resonance condition.

The loss rate can be described by two coupled rate equations of the ground-state atom number $n$ and the Rydberg state atom number $n_R$:
\begin{equation}
\frac{dn}{dt}=R-\Gamma n-(\beta_0 n_R) n
\label{eq:nloss}
\end{equation}
\begin{equation}
\frac{dn_R}{dt}=k I n-\Gamma_Rn_R-(\beta_R n) n_R,
\label{eq:rnloss}
\end{equation}
where R is the MOT capture rate, $\Gamma$ is MOT loss rate due to background collision, and $k$ is the Rydberg excitation rate. The ground-Rydberg collision results in loss rates of the ground-state atoms and the Rydberg atoms are $\beta_0$, and $\beta_R$, respectively. Given $\beta_Rn$ $\gg$ $\Gamma_R$, the steady state solution of the atom numbers are:
\begin{equation}
n_R=\frac{k I}{\beta_R+\frac{\Gamma\Gamma_R}{R}}
\label{eq:srloss}
\end{equation}
\begin{equation}
n=\frac{R}{\Gamma+\frac{I}{\beta_R/(\beta_0 k)}}.
\label{eq:snloss}
\end{equation}
Experimentally, the ground-state atom number is proportional to the observed MOT fluorescence. The $\Gamma$ was measured to $1.5~\rm s^{-1}$ using the trap rising time without Rydberg excitation (i.e., far-detuned probe laser beam). This serves as a calibration for the other loss mechanisms under investigation.

As shown in Fig.~\ref{Fig:mot-state}, the experimental
data were fitted by the trap loss model given by Eq.~\ref{eq:snloss}, where the saturation parameter $I_s=\beta_R/(\beta_0 k)$ of the trap loss (listed in Table~\ref{tab:sat-para}) can be used to quantify the power dependent loss strength. Larger losses were observed in higher $n$ states, despite the excitation rate $k$ is expected to be smaller in the higher $n$ state. However, the number of the ground-state atom
within the radius of Rydberg atom is scaled as $n^6$.
\begin{table}[htb]
\begin{tabular}{ccccccc}
\hline\hline
state & 50S & 48D & 60s & 58D & 70S & 68D \\
\hline
$I_s$(mW) & 185.1 & 54.0 & 156.7 & 54.7 & 31.1 & 10.2\\
\hline\hline
\end{tabular}
\caption{The saturation parameter $I_s$ of the trap loss for various Rydberg states was experimental determined by fitting the curve in Fig.~\ref{Fig:mot-state} based on Eq.~\ref{eq:snloss}.}\label{tab:sat-para}
\end{table}

On the other hand, the loss rates of $n$D were observed to be higher than that of $n$S. The $I_s$ of the
$n$S state is about three times higher than that of $n$D. Such differences cannot be explained by merely optical excitation rates $k$, which is estimated to only 30\% difference in the nearby $n$S and $n$D states.
Therefore, the ground-Rydberg collision part of the $I_s$ should contribute significantly to the mechanism affecting the final trap loss.
It could be understood as that, in the case of on resonance, a certain amount of Rydberg $n$S or $n$D atoms are produced, the trapped 4S ground-state atoms collide with the Rydberg atoms to escape from the trap.
In this situation, the collision cross section of the ground state ($\beta_0$) will have a direct impact on the trap loss and is relevant to the orbital angular momentum or the hyperfine manifold of the Rydberg atom. Such an enhanced collision cross section might be attributed to the anisotropic wave function of the $n$D state valence electron.

\section{Conclusion}

With a potassium hot vapor cell, we successfully demonstrate a sub-Doppler EIT spectroscopy using two-step excitation in an inverted ladder-type scheme. The transparency window with a width $<$50~MHz enable high-resolution spectroscopy and we used it to measure the energy level for the Rydberg states up to 70S with an uncertainty better than 0.03~cm$^{-1}$.
Furthermore, the good signal-to-noise ratio of EIT spectroscopy could provide optical detection of high Rydberg states, as that in normal ladder-type schemes. 

We developed a theoretical model using optical Bloch equations without the weak probe approximation, which is in an excellent agreement with our experimental results. Our experimental observations reveal that, in the vapor cell, an inhomogeneous media, the Doppler distribution of atoms plays a crucial role in narrowing the transparency window.
The observed sub-Doppler feature is typically composed of absorption and transparency features dependent on the strengths of the probe and coupling fields.
The strong correlation between the principle quantum number of the excited Rydberg states $n$ and the dephasing rates was also observed. A more pronouncing EIT in the high-$n$ states implies that the longer lifetime of the high Rydberg states reduces the dephasing rate and enhances the coherence, as predicated. 

The same excitation scheme was also performed in a cold potassium MOT to study the trap loss induced by Rydberg excitation. In comparison with the hot cells, the cold atomic ensembles with a narrow velocity distribution allows only two-photon excitation. The mechanisms behind the surprisingly large trap loss were studied in various Rydberg states and detuning conditions. For the case of far-detuning from the intermediate state, the loss is found to be resulted from the perturbation to the levels related to the laser tapping scheme, due to the ac Stark shift. For the case of that both coupling and probe lasers are nearly on resonances (two-step resonance), we observed that the interaction cross section between the Rydberg atom and the ground state rapidly grows as the principle quantum number $n$ increases. In addition, the $n$D state Rydberg atom ($l=2$) exhibits a larger loss of the ground-state atom than that of the $n$S state. It could be attributed to the valence electron's higher orbital angular momentum or its hyperfine manifold. Our experiment provides a detailed study on the potassium Rydberg excitation for the future quantum technology applications using the Rydberg-dressed potassium, especially the heteronuclear qubits.

\begin{acknowledgments}
Research was supported by the Center for Quantum Technology from the Featured Areas Research Center Program within the framework of the Higher Education Sprout Project by the Ministry of Education (MOE) in Taiwan, and the Ministry of science and Technology (MOST) under Grant 106-2112-M-007-021-MY3 and 105-2112-M-007-027-MY3. 
\end{acknowledgments}

%

%

\end{document}